\documentclass[prl,showpacs,preprintnumbers,amsmath,amssymb,letterpaper,twocolumn]{revtex4}

\usepackage{graphicx}
\usepackage{dcolumn}
\usepackage{bm}
\usepackage{epstopdf}
\usepackage{latexsym}
\usepackage{epsfig}
\usepackage{verbatim}
\usepackage{hyperref}
\usepackage{color}

\newcommand{\be}{\begin{equation}}
\newcommand{\ee}{\end{equation}}
\newcommand{\bea}{\begin{eqnarray}}
\newcommand{\eea}{\end{eqnarray}}

\begin{document}

\title{Exchange Field-Mediated Magnetoresistance\\ in the Correlated Insulator Phase of Be Films }

\author{T.J Liu, J. C. Prestigiacomo, Y.M. Xiong, and P.W. Adams}
\affiliation{Department of Physics and Astronomy, Louisiana State 
University, Baton Rouge,
Louisiana 70803, USA}

\date{\today}

\begin{abstract}
We present a study of the proximity effect between a ferromagnet and a paramagnetic metal of varying disorder.  Thin beryllium films are deposited onto a 5 nm-thick layer of the ferromagnetic insulator EuS.  This bilayer arrangement induces an exchange field, $H_{ex}$, of a few tesla in low resistance Be films with sheet resistance $R\ll R_Q$, where $R_Q=h/e^2$ is the quantum resistance.   We show that $H_{ex}$ survives in very high resistance films and, in fact, appears to be relatively insensitive to the Be disorder.  We exploit this fact to produce a giant low-field magnetoresistance in the correlated insulator phase of Be films with $R\gg R_Q$.    
\end{abstract}


\maketitle

It is now well established that the interfacial region between two materials of differing order parameters can have subtle and even counter-intuitive properties.   Indeed, some well known examples of such ``proximity effect'' systems have important technological applications, as well.  For example, an additional exchange field can be induced into a ferromagnetic (FM) film by placing it in contact with an appropriate antiferromagnet.  The resulting ``exchange bias'' shifts the FM hysteresis loop away from zero field \cite{Ex-bias}.  Although the microscopic mechanism of the exchange bias is not well understood, it is, nevertheless, an important component of many magnetic data storage technologies.  If the antiferromagnet is replaced with a superconductor (SC), a SC order parameter can be induced in the FM component with a complementary exchange field induced in the SC component. In fact, recent studies of FM/SC hetereostuctures \cite{Buzdin1,Beasley,Efetov,Kontos} have shown that not only can Cooper pairs exist in the FM layer, but the SC order parameter oscillates in sign on the FM side of the structure.  This effect is the basis of the $\pi$ Josephson junction \cite{Buzdin2,Bannykh}.  Similarly, trilayer configurations such as FM/SC/FM can be used to produce superconducting spin-switch devices \cite{Velthuis,Okamoto}. In this Letter we present a study of the proximity effect between a FM insulator and a disordered paramagnetic (PM) film.   This relatively simple system, which only has a single order parameter, gives one the opportunity to investigate what roles disorder and $e-e$ correlations play in establishing an exchange field in the PM layer.   As we show below, exchange fields with magnitudes much greater than the saturation magnetization field of the FM can routinely be induced in the PM films.  These exchange fields are insensitive to disorder and, in fact, can persist well into the highly correlated variable-range-hopping regime of the PM.  In this latter limit, the exchange field can modulate the internal field of the PM layer in such a manner as to produce extremely large low-field magnetoresistances (MR).    

Here we have chosen to study the proximity effect between the ferromagnetic insulator EuS and ultra thin beryllium films.  At low temperatures the EuS underlayer is highly insulating, therefore the transport currents are confined to the Be layer.  Beryllium forms smooth, dense, non-granular
films when deposited via electron-beam evaporation.   This non-granular morphology is crucial in that it assures one that the measured resistance reflects $e-e$
correlation effects and {\em not} grain charging effects \cite{Be}.  Beryllium has the additional advantage of having a superconducting phase in low resistance films \cite{SC-Be}.  This phase is suppressed in films with sheet resistances $R>10$ k$\Omega$/sq. Films with $R\gtrsim R_Q$, where $R_Q=h/e^2$ is the quantum resistance exhibit a low temperature correlated insulator phase associated with modified variable range hopping and the opening of a 2D Efros-Shklovskii Coulomb gap \cite{ES,CG}.  Interestingly, applying a magnetic field fills the Coulomb gap, thereby producing extremely large decreases in resistance at low temperatures \cite{Be-GMR}.  In the high field limit, the films enter a ``quantum metal'' phase, in which the MR saturates at $R\sim R_Q$ \cite{QM}.  Thus, Be offers almost ideal opportunity to study the exchange field from the weakly to the strongly interacting limits. 

 The bilayers were formed by first depositing a 5 nm thick EuS on fire-polished glass at 84 K. Subsequently, a Be layer with thickness ranging from
1.8 - 2.3 nm was deposited directly on top of the EuS.   The EuS layer did not conduct, but the Be layers had low temperature resistances ranging from
$R \approx 1$ k$\Omega$/sq to 10 M$\Omega$/sq.   The evaporations were made in
a $4\times10^{-7}$ Torr vacuum at a rate $\sim0.1$ nm/s using e-beam deposition.  The film conductances were measured using
a standard four probe ac lock-in technique for low resistance samples, but dc I-V's were used in high R films.  MR measurements were made using a Quantum Design PPMS and a dilution refrigerator, both utilizing a 9 T superconducting solenoid.

\begin{figure}
\begin{flushleft}
\includegraphics[width=.44\textwidth]{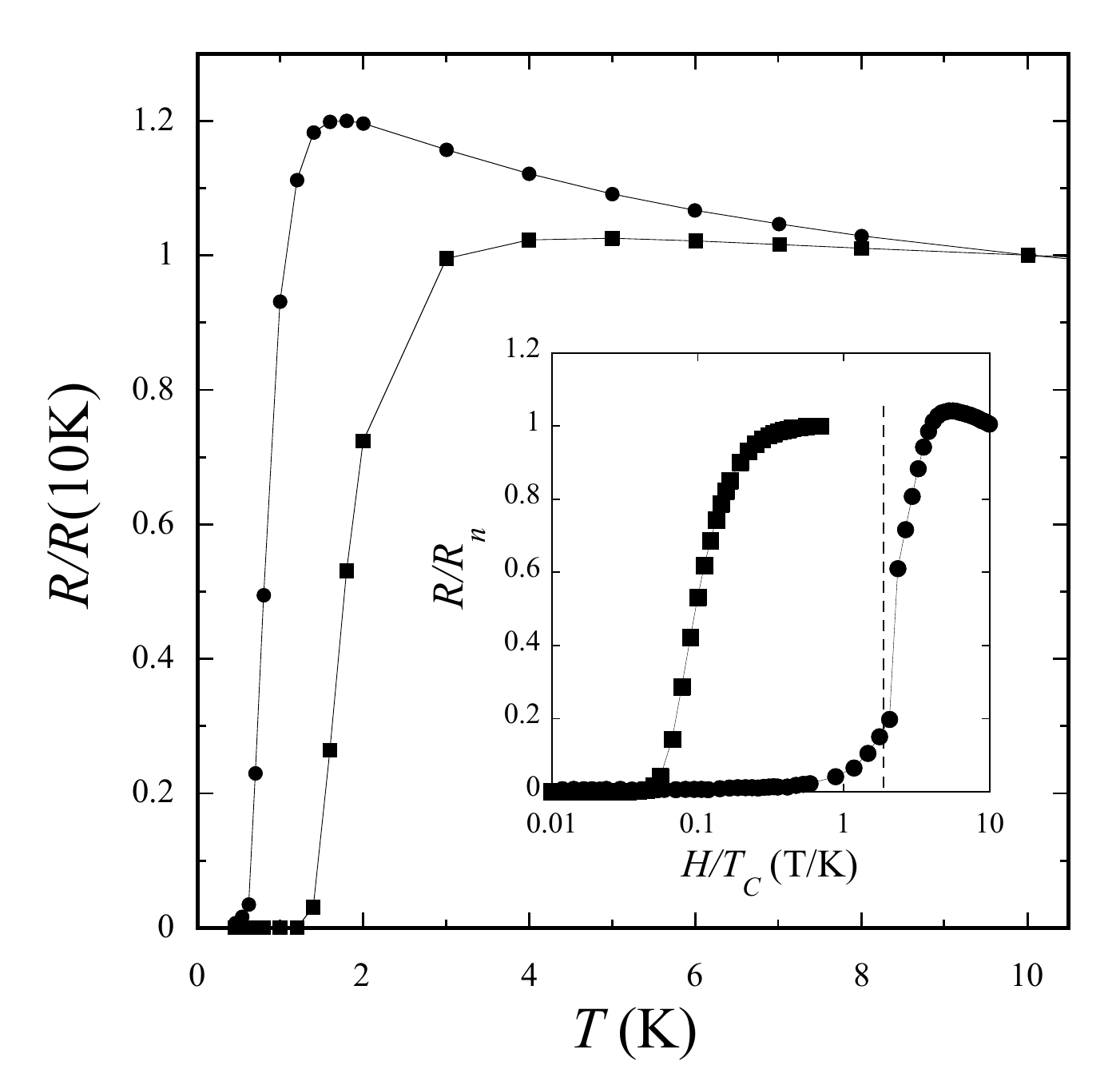}\end{flushleft}
\caption{\label{TcHc} Temperature dependence of a 2.5 nm-thick Be film on glass (circles) and a 2.5 nm-thick Be film on a 5 nm-thick EuS film (squares).  Inset: film resistance as a function of reduced parallel field at 460 mK.  The vertical dashed line is the expected Clogston-Chandresekhar critical field.}
\end{figure}

Previous spin-polarized tunneling studies of the magnetic proximity effect in EuS/Al bilayers show that an exchange field, $H_{ex}$, of several tesla can be induced in Al films at low temperatures \cite{Tedrow,Hao,Xia}.  In the presence of an applied field, $H_{app}$, the total internal field in the Al layer is $H_{int}=H_{app}+H_{ex}$.  Interestingly, in the EuS/Al system $H_{ex}$ is not static, but increases as $\ln (H_{app})$ in parallel applied fields between 0.01~T and 2~T \cite{Xiong}.  We believe that $H_{ex}$ behaves similarly in the EuS/Be layers of this study.  Interestingly, this effect is not an artifact of domain alignment in the EuS under-layer.  Direct measurements of the magnetization of the EuS/Al bilayers revealed a very sharp in-plane magnetization loop with a 2 K coercive field of $\sim5$x$10^{-3}$ T \cite{Xiong,Note}.   Thus the magnetization of the EuS was saturated over most of the field range where the $\ln (H_{app})$ behavior was observed.  Recent measurements of exchange field effects in FM/SC/FM trilayers have also shown that MR of these systems cannot be explained in terms of the magnetization behavior of the FM layers \cite{Velthuis,Okamoto}.  Instead the MR arises from the interplay between the applied field and the evanescent tail of the exchange field.

  Shown in the main panel of Fig.~\ref{TcHc} is the superconducting transition for a 2.3 nm-thick Be film on glass and a comparable Be film on EuS.  In each case $R\sim1$ k$\Omega$/sq.  Unexpectedly, the transition temperature $T_c$ of the bilayers was higher than that of the pristine Be films deposited on glass.  In the inset we plot the corresponding parallel critical field of the samples as measured at 460 mK.  The field scale has been normalized by the respective transition temperature of the two samples.  Since the thickness of the Be layers was much less than the coherence length, $\xi\sim30$ nm, the critical fields in the inset are Zeeman limited \cite{Tedrow}.  In terms of $T_c$, the expected critical field is given by Clogston-Chandresekhar relation $H^{cc}_c=1.86\times T_c$ \cite{CC,Tinkham}.  Note that for Be on glass $T_c=0.9$ K and $H_{c||}=2.3$ T, in reasonable agreement with $H^{cc}_c$, see dashed line in the inset of Fig.~\ref{TcHc}.  In contrast, the EuS/Be sample has a $T_c=1.8$ K, but its critical field, $H_{c||}=0.1$ T, is 20 times less than that of the pristine Be film!   If there were no exchange field in the Be component of the EuS/Be bilayer, then we would expect $H_{c||}\approx4.6$ T. Therefore, we can surmise that $H_{ex}\sim 4.5$ T at the critical field transition of the bilayer.  

\begin{figure}
\begin{flushleft}
\includegraphics[width=.44\textwidth]{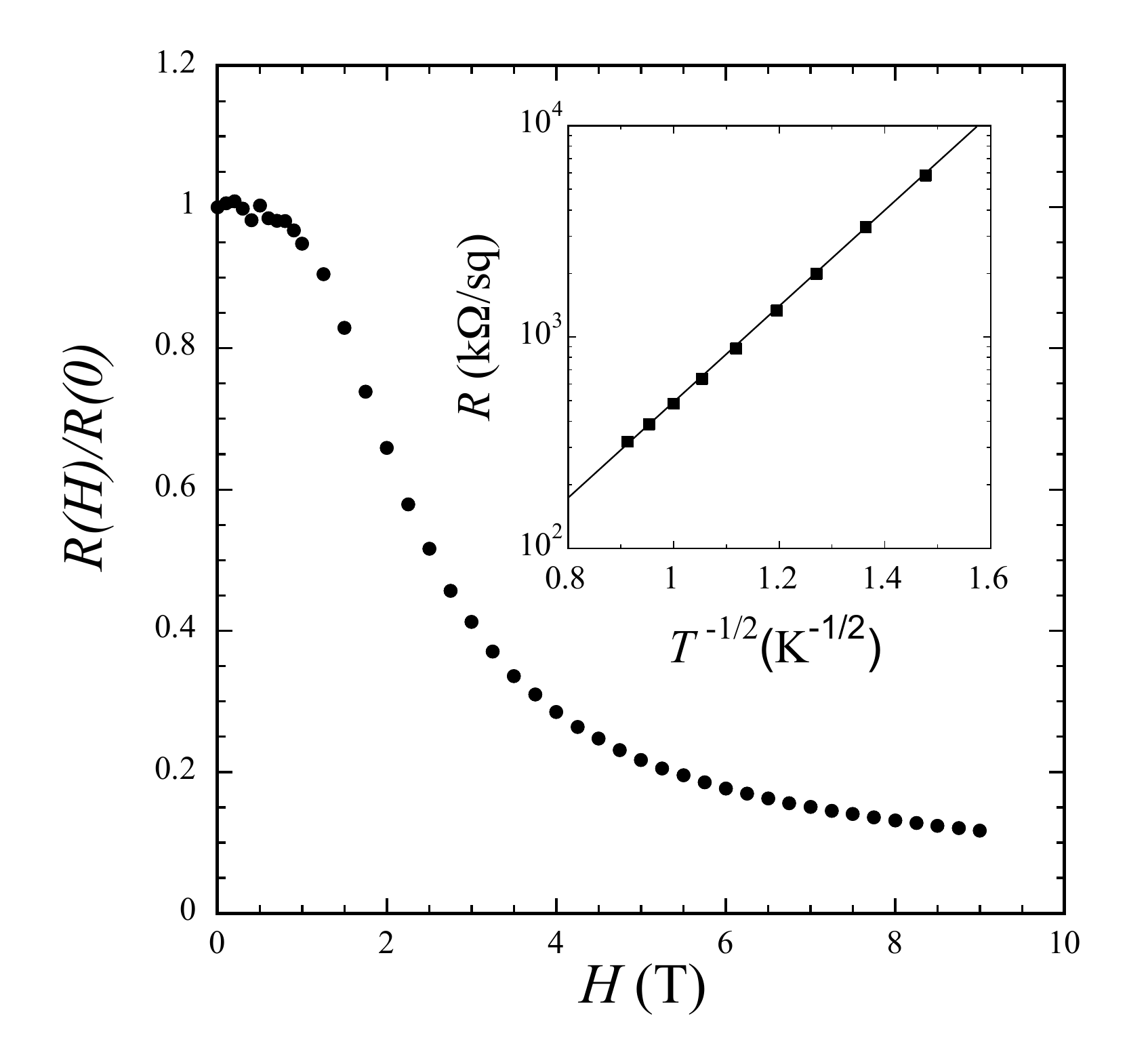}\end{flushleft}
\caption{\label{PristineMR} Magnetoresistance of a pristine 1.6 nm-thick Be film on glass at 460 mK.  The film had a zero-field sheet resistance of $R=230\times R_Q$.  Inset: zero-field temperature dependence of the film showing modified variable range hopping behavior. }
\end{figure}

Most studies of the exchange field in FM/PM systems have been made in the superconducting phase of a low atomic mass PM.  The reason for this is that the Zeeman coupling to the superconducting quasiparticles gives one a direct probe of the induced exchange field via tunneling density of states, but only if the spin-orbit scattering rate is low \cite{,Tedrow,Hao}.  Under special circumstances, one can also use a Cooper pair resonance to probe $H_{ex}$ in the Zeeman-limited normal state of the PM, again via tunneling density of states \cite{Xiong,Xiong2}.  In each case, however, the behavior of $H_{ex}$ is explored in the context of the superconducting correlations in relatively low resistance films. Indeed, the role of such correlations on the manifestation of $H_{ex}$ is unclear. Here we exploit the very large low-temperature MR of high resistance Be films \cite{Be-GMR} to probe $H_{ex}$ in an unexplored region of parameter space.  

Shown in Fig.~\ref{PristineMR} is the MR of a pristine $1.8$ nm-thick Be film deposited on glass.   At 460 mK the sheet resistance of this film is well above $R_Q$.   The film's transport is of the modified variable range hopping form as is evident in the inset of Fig.~\ref{PristineMR}.  Note that the parallel field MR is non-perturbative, with the resistance falling by a factor of 10 in a field of 9 T.  The orbital contributions to the MR are negligible in parallel field. Consequently, the MR is completely driven by the Zeeman coupling to the conduction electrons.  This MR is believed to be intrinsic to the correlated insulator phase of the Be films \cite{Be-GMR}.    If a substantial $H_{ex}$ can be established in this phase, then it will certainly affect the MR via its contribution to the Zeeman splitting.  So it is useful to compare and contrast the pristine MR curve in Fig.~\ref{PristineMR} with what we obtain from similar measurements on EuS/Be bilayers.

\begin{figure}
\begin{flushleft}
\includegraphics[width=.44\textwidth]{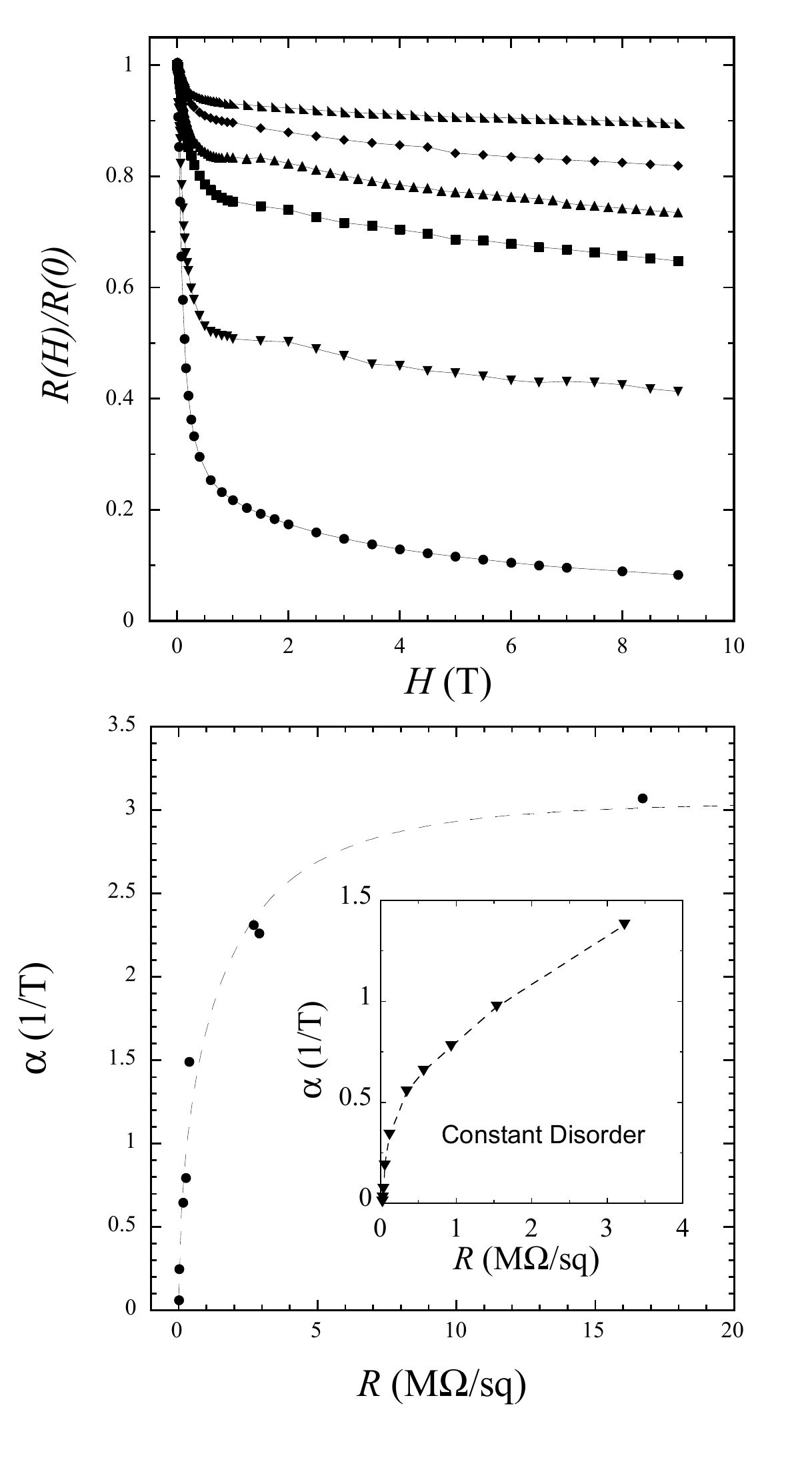}\end{flushleft}
\caption{\label{BilayerMR} Upper Panel: Parallel field MR curves for a set of EuS/Be bilayers with varying Be resistance.  The data were taken in parallel magnetic field at 460 mK. Moving from the upper trace to the lower one, the zero-field sheet resistances are 0.02, 0.04, 0.16, 0.26, 2.4, and 17.0 M$\Omega$/sq.  Lower Panel: The relative field sensitivity $\alpha$ as a function of the zero-field 460 mK sheet resistance for samples of varying resistance, including some not displayed in the upper panel. The dashed line is provided as a guide to the eye.  Inset: field sensitivity of the 0.16 M$\Omega$/sq sample as a function of $R(T)$.  Here the temperature was varied but the disorder remained constant.  The dashed line is a guide to the eye.}
\end{figure}

Shown in the upper panel of Fig.~\ref{BilayerMR} is the MR of several EuS/Be bilayers of varying Be thickness and corresponding sheet resistance.  All of these data were taken in parallel field at 460 mK.   The EuS thickness of each sample was 5 nm. In contrast to Fig.~\ref{PristineMR}, the bilayers exhibit an extremely large low-field negative MR.  Indeed, the resistance of the 17 M$\Omega$/sq falls a factor of two in an applied field of only $\sim0.2$ T. In fact, the MR of the bilayers looks vaguely similar to that of their pristine Be counterparts except for the fact that the field scale has been greatly compressed.  We believe this is due to the rapid emergence of $H_{ex}$ with applied field.  Spin-resolved tunneling measurements in EuS/Al  bilayers show that the magnitude of $H_{ex}$ increases logarithmically with applied field up to about 2 T, at which point $H_{ex}$ reaches its saturation value \cite{Xiong}.  The knee in the MR curves corresponds well with this saturation point.  We do not have a direct measure of the actual value of the saturated exchange field, but it can be as high as 10 T in EuS/Al bilayers.  The MR beyond the knee is attributable to applied field alone. 

\begin{figure}
\begin{flushleft}
\includegraphics[width=.44\textwidth]{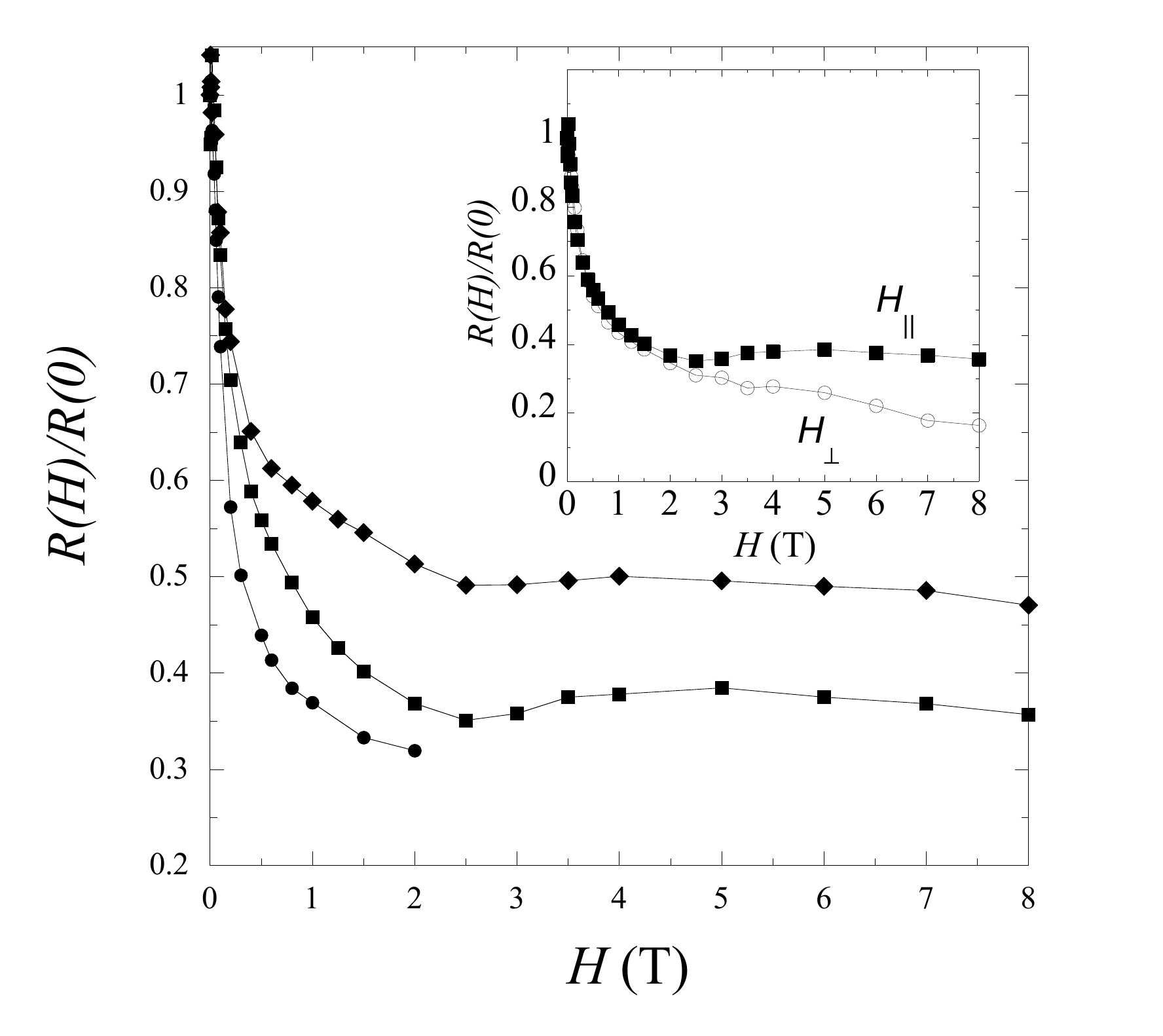}\end{flushleft}
\caption{\label{MR-T} Parallel field MR traces at 52 mK (circles), 80 mK (squares) and 130 mK (diamonds) for the 0.16 M$\Omega$/sq bilayer of Fig.~\ref{BilayerMR}.  Inset: Comparison of the relative MR in parallel (squares) and perpendicular (circles) applied field at 80 mK.  }
\end{figure}

In order to quantify the strength of the low-field MR, we define the relative field sensitivity of the resistance as $\alpha=\left|\Delta R\right|/\left|(R(0)\Delta H)\right|$ for data below $0.2$ T.   Shown in the lower panel of Fig.~\ref{BilayerMR} are the values of $\alpha$ at 460 mK for the corresponding samples of the upper panel. In pristine Be films the overall size of the MR increases with increasing resistance \cite{Be-GMR,QM} but extends over a field range of $\sim10$ T.  The MR of the bilayers also grows with increasing disorder as is evident in Fig.~\ref{BilayerMR}. These data indicate that $H_{ex}$ is relatively insensitive to film disorder, even well into the correlated insulator regime.   This suggests that the spin flip scattering rate in the Be remains low in films with $R\gg R_Q$, otherwise spin relaxation events would wash out the exchange field.  The resistance of the films in Fig.~\ref{BilayerMR}  is in large part controlled by the Be thickness, with thinner films having a higher low-temperature R.  But $H_{ex}$ itself is known to increase in proportion to the inverse of the PM film thickness \cite{Hao}.  Thus, the large values of $\alpha$ in our highest resistance samples can, in part, be attributed to the fact that the Be component of those bilayers was thinner.  This, however, was probably not the dominant factor in the behavior of the lower panel of Fig.~\ref{BilayerMR}  since the film thicknesses used in this study varied less than 30\%.     

For our most resistive films $\alpha\approx2.5$ T$^{-1}$, which is somewhat larger than what has been reported in other high-MR systems.  For instance, $\alpha\sim0.6$ T$^{-1}$ for Fe/Cr superlattices \cite{FeCr-GMR}, $0.7$ T$^{-1}$ for silver chalcogenides \cite{AgTe-GMR}, and $1.2$ T$^{-1}$ for LaSb$_2$ \cite{LaSb2}.   In the inset of Fig.~\ref{BilayerMR} we show the temperature dependence of $\alpha$ for the $0.16$ M$\Omega$/sq sample in the upper panel of the figure.  In this case the disorder remains constant, but the magnitude of $\alpha$ increases with decreasing temperature.   This effect is clearly evident in the MR curves in Fig.~\ref{MR-T}.  When the transport measurements on the $0.16$ M$\Omega$/sq sample of Fig.~\ref{BilayerMR} are extended down to 50 mK, the relatively modest 20\% MR at 460 mK  increases to more than 300\%.  Lowering the temperature amplifies the effect of $H_{ex}$ due to fact that the sample is deeper into the correlated insulator phase at low temperatures.  Finally, in the inset of Fig.~\ref{MR-T}  we compare the parallel and perpendicular MR curves of the sample at 50 mK.   Note, that, below 2 T, the MR traces are almost identical in each field direction.  This, of course, is what is expected if the MR in this region is dominated by an exchange field whose magnitude is an isotropic function of the applied field.    

In summary, the emergence of a large, negative low-field MR in high resistance EuS/Be bilayers offers proof that a significant exchange field can be induced in the two-dimensional correlated insulator phase of a disordered paramagnet.  Both the magnitude and field dependence of $H_{ex}$ appear to be comparable to what is observed in low disorder superconducting bilayers.  Indeed, the multifold MR of our highest resistance samples can be attributed to the rapid increase in the magnitude of $H_{ex}$ upon the application of relatively modest external magnetic fields.  More generally, any property of the PM layer that is a function of the conduction electron Zeeman-splitting will be similarly affected by the onset of the exchange field.

\acknowledgments

We gratefully acknowledge enlightening discussions with Ilya Vekhter and John DiTusa.  This work was supported by the DOE under
Grant No.\ DE-FG02-07ER46420.


\begin{thebibliography}{30}
\bibitem{Note} The 2 K M-H magnetization loops of 5.0-nm/2.5-nm EuS/Be bilayers were essentially identical to their EuS/Al bilayers counterparts.
\end{thebibliography}
\end{document}